\def\slash#1{\setbox0=\hbox{$#1$}#1\hskip-\wd0\dimen0=5pt\advance
\dimen0 by-\ht0\advance\dimen0 by\dp0\lower0.5\dimen0\hbox
to\wd0{\hss\sl/\/\hss}}
\def\Black{}
\def\Blue{}
\def\Brown{}

\documentstyle[aps,epsf,prl]{revtex}
\newcommand{\be}{\begin{equation}}
\newcommand{\ee}{\end{equation}}

\begin{document}

\begin{titlepage}
\null
\begin{center}
\Large\bf \Brown The semileptonic $B\to \pi$ decay in a
Constituent Quark-Meson model. \Black
\end{center}
\vspace{0.5cm}

\begin{center}
\begin{large}
A. Deandrea\\
\end{large}
\vspace{0.3cm}
Institut f\"ur Theoretische Physik, Universit\"at Heidelberg,\\
Philosophenweg 19, D-69120 Heidelberg, Deutschland\\
\vspace{0.5cm}
\begin{large}
R. Gatto\\
\end{large}
\vspace{0.3cm}
D\'epartement de Physique Th\'eorique, Universit\'e de Gen\`eve,\\
24 quai E.-Ansermet, CH-1211 Gen\`eve 4, Suisse\\
\vspace{0.5cm}
\begin{large}
G. Nardulli and A. D. Polosa\\
\end{large}
\vspace{0.3cm}
Dipartimento di Fisica, Universit\`a di Bari and INFN Bari,\\
via Amendola 173, I-70126 Bari, Italia
\end{center}

\vspace{0.5cm}

\begin{center}
\begin{large}
\Brown
{\bf Abstract}\\[0.5cm]\Black
\end{large}
\parbox{14cm}{We evaluate the form factors describing the exclusive decay
$B\to\pi l \nu$ by using a Constituent Quark-Meson model based on
an effective quark-meson Lagrangian (CQM). The model allows for an
expansion in the pion momenta and we consider terms up to the
first order in the pion field derivatives. We compute the leading
terms in the soft pion limit and consider corrections to this limit.}
\end{center}

\vspace{1.0cm}
\noindent
\Blue
\Black
\vfil
\noindent
\Brown
BARI-TH/99-341\\
UGVA-DPT 1999/06-1045\\
July 1999
\Black
\end{titlepage}

\setcounter{page}{1}
\twocolumn

\preprint{BARI-TH/99-341\\
UGVA-DPT 1999/06-1045}
\title{The semileptonic $B\to \pi$ decay in a
Constituent Quark-Meson model.}
\author{A. Deandrea}
\address{Institut f\"ur Theoretische Physik, Universit\"at Heidelberg,
Philosophenweg 19, D-69120 Heidelberg, Deutschland}
\author{R. Gatto}
\address{D\'epartement de Physique Th\'eorique, Universit\'e de Gen\`eve,
24 quai E.-Ansermet, CH-1211 Gen\`eve 4, Suisse}
\author{G. Nardulli and A. D. Polosa}
\address{Dipartimento di Fisica, Universit\`a di Bari and INFN Bari,
via Amendola 173, I-70126 Bari, Italia}
\date{July 1999}
\maketitle
\begin{abstract}
We evaluate the form factors describing the exclusive decay
$B\to\pi l \nu$ by using a Constituent Quark-Meson model based on
an effective quark-meson Lagrangian (CQM). The model allows for an
expansion in the pion momenta and we consider terms up to the
first order in the pion field derivatives. We compute the leading
terms in the soft pion limit and consider corrections to this limit.
\end{abstract}

\section{Introduction}

The investigation of the semileptonic decay $B\to\pi l \nu$ is
relevant for the extraction of the Kobayashi-Maskawa matrix element
$V_{ub}$. The analysis of this exclusive
decay mode would offer a method alternative to the inclusive
semileptonic $B$ decay for the study of the  $b \to u$
transition. A precise measurement of this process
is one of the main aims of the future $B$-factories.

On the theoretical side this decay process  has received large
attention in literature (see for example the review in
\cite{rep,rev}) since it offers an example of heavy-to-light quark
transition computable by the presently available theoretical
methods.

The present
note is devoted to the study of the $B \to \pi l \nu$ decay mode
in the framework of the Constituent-Quark-Meson (CQM) model
\cite{noi}. In this model the transition amplitudes are
evaluated by computing  diagrams in which heavy and light mesons
are attached to quark loops. Moreover, the light chiral symmetry 
restrictions and the
heavy quark spin-flavour symmetry dictated by the
Heavy-Quark-Effective-Theory (HQET) are both implemented. The advantage
of such a description is the reduced number of free parameters
with respect to an effective Lagrangian at the meson level with no
dynamical assumptions\cite{semil}.

A short glossary, useful to go through the results reported here,
is in order. We call $H$  the field representing the low-lying
heavy meson doublet $(0^-,1^-)$~\cite{falk}, $Z_H$  the heavy
field renormalization constant, induced by loop effects and
$\Delta_H$ the difference between the $H$ meson doublet mass and
the mass of the constituent heavy quark. $\Delta_H$ is an
adjustable parameter of the model and we restrict to
$\Delta_H=0.4\pm 0.1$~GeV since only this range of values allows
for a good phenomenology of semileptonic weak decays (for a
discussion see \cite{noi}). For the definition of the model, it is
important to fix the regularization procedure allowing to
calculate explicitly the quark loop integrals. We use the
Schwinger proper time regularization method, assuming, as
ultraviolet (UV) and infrared (IR) cut-off, $\Lambda\simeq 1.25$
GeV and $\mu\simeq 0.3$ GeV respectively. Another parameter is
the constituent light quark mass $m$ that we have fixed in
\cite{noi} to the value: $m=0.3$~GeV for $u$ and $d$ flavours.

\section{$B \to \pi$ form factors}
We consider the weak current matrix
element for the semileptonic $B\to \pi$ transition which is given
by ($q=p-q_\pi$):
\begin{eqnarray}
\langle \pi(q_\pi)|V^{\mu}(q)|B(p)\rangle &=&
\left[ (p+q_\pi)^{\mu}+\frac{m_{\pi}^2-m_B^2}{q^2}q^\mu \right]\;
F_1(q^2)\nonumber \\
&-& \left[ \frac{m_{\pi}^2-m_B^2}{q^2}q^{\mu} \right] \; F_0(q^2) \; .
\end{eqnarray}
with $F_1(0)=F_0(0)$. The calculation of the semileptonic process
proceeds through the evaluation of the diagrams in Fig.~1,2 and 3.
Fig.~1 gives rise to a non derivative coupling. Fig.~2a and Fig.~2b
are polar diagrams where the pion is introduced through a
derivative interaction term: in Fig.~2a the intermediate particle
is the vector meson particle belonging to the $H$ heavy meson
multiplet;  in Fig.~2b the intermediate particle is the scalar
($J^P=0^+$) meson particle belonging to the positive parity $S$
heavy meson multiplet (this multiplet is built similarly to $H$
and contains also an axial vector meson $J^P=1^+$ state). 
The diagram in Fig.~2b represents only a correction
to the chiral symmetry limit. 
To obtain the contribution of  Fig.~1, an expansion
of the chiral rotated light quark field, $\chi$, up to the first
order in $\pi$ is needed~\cite{noi,ebert}. In the case of Fig.~2
and Fig.~3 the same expansion is truncated at the zero-th order.
The $\chi$ field, as defined in~\cite{ebert}, is given by
$\chi=\xi q$ being $q$ the usual spinor field describing the light
degrees of freedom and $\xi=e^{i\pi/f_\pi}$, with $f_\pi=130~{\rm
MeV}$.

The diagram in Fig.~1 produces a result proportional to the
leptonic $B$-decay constant; its predictions are expected to be
valid at small pion momenta, near the zero recoil point: $q^2_0
\simeq (m_B-m_\pi)^2$. One obtains, from this non-derivative (ND)
coupling, the contributions:
\begin{eqnarray}
F_{0}^{\rm ND}&=&\frac{f_B}{f_\pi} \\ F_{1}^{\rm
ND}&=&\frac{f_B}{2f_\pi}~~,
\label{catr}
\end{eqnarray}
where $f_B$ is the $B$ leptonic decay constant. The CQM-model
evaluation of $f_B$ is given in \cite{noi}. Neglecting a smooth
logarithmic dependence, the heavy meson mass dependence of $f_B$,
as predicted by the HQET, is as follows:
\begin{equation}
f_B=\frac{{\hat F}}{\sqrt{m_B}}
\label{effe}
\end{equation}
where $\hat F$ parametrises the leading term in the decay constant
$f_B$ (see for example \cite{rep}).
Next we consider the polar diagrams in Fig.~2. First of all let us
consider the diagram in Fig.~2(a) which gives a contribution
proportional to $g \hat F$, where $g$ is the $HH\pi$-strong
coupling constant. One
obtains the following contribution to $F_1$~\cite{rep,wise}:
\begin{equation}
F_{1}^{\rm Pol}(q^2)= \frac{\hat{F}g}{f_{\pi} \sqrt{m_B}}
\frac{1}{1-q^2/m_{B^*}^2} \; .
\label{funopol}
\end{equation}
where $g$ is the $HH\pi$-strong coupling constant evaluated
in \cite{noi} (see also \cite{nasum} for an evaluation in the framework
of the QCD sum-rule approach and \cite{lagpi} in the framework of
the effective meson Lagrangian approach).
The diagram in Fig.~2(b) contributes to the other form factor:
\begin{equation}
F_{0}^{\rm Pol}(q^2)=\frac{1}{m_B^2-m_\pi^2}\left(
\frac{h m_\pi \sqrt{m_B} {\hat F}^+}{f_\pi}\right)\frac{1}{1-q^2/m_{B^{**}}^2}~~,
\label{fzeropol}
\end{equation}
where $h$ is the $HS\pi$-strong coupling constant evaluated in
\cite{noi}, $B^{**}$ is the $0^+$ state in the $S$ multiplet
$(0^+,1^+)$ and ${\hat F}^+$ is the $B^{**}$ leptonic decay
constant analogous to ${\hat F}$. All the coupling constants
appearing in these equations can be computed in the CQM. The polar
results  (\ref{funopol}) and (\ref{fzeropol}) should be reliable
near the poles, i.e. again for $q^2$ large, around $q^2_0$, the
zero recoil point. We shall discuss below a procedure to
extrapolate these results to lower $q^2$ values. For future
reference we quote here the values of  the expressions
(\ref{funopol}) and (\ref{fzeropol}) at $q^2=0$:
\begin{eqnarray}
F_{1}^{\rm Pol}(0)&=& 0.52 \pm 0.01\\ F_{0}^{\rm Pol}(0)&=&0.012\pm
0.001 \; .
\end{eqnarray}
They are obtained with the values $\hat F= 0.34\pm 0.02,~\hat F^+=
0.24\pm 0.03,~g= 0.46\pm 0.04,h=- 0.76\pm 0.13 $ \cite{noi}.

The results obtained so far are
not new; they have been obtained by several groups and our
contribution consists here only in the calculation, within the
CQM-model, of the various parameters appearing in the previous
equations. The model however allows to consider a new
contribution, depicted in Fig.~3: it differs from the 
ND (non-derivative) term
since it is derivative and from the polar term because it does not
contain couplings to resonances. The current
directly couples to the quark in this case rather than to the heavy
meson as in the polar contribution. Both the polar and the new ``direct''
contributions can be reliably calculated only at large $q^2$, note however 
that differently from (7) and (8) for the polar terms, the relation 
$F_1(0)=F_0(0)$ will be automatically satisfied by the new contributions
to be described in the following. We compute them by a
straightforward application of the basic rules of the CQM. As the current
will transform a heavy meson into a light one in this case, 
we need the interaction
of the pion with light quarks. We recall here the corresponding Lagrangian. 
The term relevant for the calculation will be the one containing an odd 
number of pions. The following term defines the Feynman rule we follow to 
insert the pion in our CQM diagram:
\begin{equation}
{\cal L}=\bar{\chi}(iD^{\mu}\gamma_\mu+{\cal A}^\mu \gamma_\mu
\gamma_5)\chi-m\bar{\chi}\chi+\frac{f_\pi^2}{8}\partial_\mu \Sigma^\dagger
\partial^\mu \Sigma \; .
\end{equation}
Apart from the mass term, ${\cal L}$, is chiral invariant. Here
$\chi$ is the chiral rotated light quark field quoted above,
$\Sigma=\xi^2$ and $\pi$ is the $3\times 3$ matrix representing the
flavour $SU(3)$ octet of pseudoscalar mesons. Moreover,
$D_\mu=\partial_\mu-i{\cal V}_\mu$ and:
\begin{eqnarray}
{\cal V}^\mu &=& \frac{1}{2}(\xi^\dagger \partial^\mu \xi+\xi
\partial^\mu \xi^\dagger) \label{vi}\\
{\cal A}^\mu &=& \frac{1}{2}(\xi^\dagger \partial^\mu \xi-\xi
\partial^\mu \xi^\dagger) \; ,
\label{aa}
\end{eqnarray}
where (\ref{vi}) generates couplings of an even number of mesons to
the $\bar{\chi}\chi$ pair, while (\ref{aa}) gives an odd number
of $\pi$ fields.

An explicit calculation of the mentioned diagram gives:
\begin{eqnarray}
F_{1}^{\rm Dir}(q^2)&=&\frac{2}{f_{\pi}}\sqrt{\frac{Z_H}{m_H}}
\left[q_\pi (C - m A) + m_H \;B\right]
\label{primoder}\\
F_{0}^{\rm Dir}(q^2)&=&\frac{2}{f_{\pi}}\sqrt{\frac{Z_H}{m_H}}
\left[\left(1-\frac{q^2}{m_{\pi}^2-m_B^2}\right) q_\pi (C-m A) \right.
\nonumber \\
&+& \left. m_H \;B \left(1+\frac{q^2}{m_{\pi}^2-m_B^2}\right)\right] .
\label{secondoder}
\end{eqnarray}
where:
\begin{eqnarray}
A&=&\frac{1}{2 q_\pi}(I_3(\Delta_H-q_\pi)-I_3(\Delta_H))\\
B&=& m A - m^2 Z(\Delta_H)\\
C&=&\frac{1}{2 q_\pi} (\Delta_H I_3(\Delta_H)
- (\Delta_H-q_\pi) I_3(\Delta_H-q_\pi))\; ,
\end{eqnarray}
and
\begin{eqnarray}
Z(\Delta) &=& \frac{N_c}{16\pi^{3/2}} \int_{1/\Lambda^2}^{1/\mu^2}
\frac{ds}{s^{1/2}}  e^{-s m^2} \nonumber \\
&\times& \int_{0}^{1}dx e^{s \Delta^2(x)}
[1+{\mathrm {erf}}  (\Delta (x)\sqrt{s})] 
\label{zeta}
\end{eqnarray}
\begin{eqnarray}
I_3(\Delta) &=&{N_c \over {16\,{{\pi }^{{3/2}}}}}
\int_{1/{{\Lambda}^2}}^{1/{{\mu }^2}} {ds \over {s^{3/2}}}
\; e^{- s( {m^2} - {{\Delta }^2} ) }\nonumber \\
&\times& \left( 1 + {\mathrm {erf}} (\Delta\sqrt{s}) \right)~,
\label{kk}
\end{eqnarray}
where $q^{\mu}_\pi=(q_\pi,0,0,q_\pi)$ is the pion 4-momentum
($m_\pi\to 0$) and $\Delta (x)= \Delta - x q_\pi$. We observe that
the soft pion limit of the previous expression brings $Z(\Delta)
\to I_4(\Delta)$ defined in \cite{noi}. Notice that in Eqs.
(\ref{primoder},\ref{secondoder}) the $q^2$ dependence arises
because of $q_\pi=(m_B^2-q^2)/2m_B$ and can be computed
numerically. We find:
\begin{equation}
F_{1}^{\rm Dir}(q^2=0)=F_{0}^{\rm Dir}(q^2=0)=0.13 \pm 0.05~~,
\end{equation}
The error in the numerical evaluations  is due to the variation of
$\Delta_H$ in the range of values $0.3-0.5$. The contribution of
the direct diagram in Fig.~3 is an appreciable $10\%$ to $30\%$ 
correction, depending on the region in $q^2$.

Since the three contributions to the form factors $F_0$ and $F_1$
are independent, we can sum them up, with the result:
\begin{equation}
\hat{F_j}(q^2)=\frac{f_B}{(j+1)f_\pi}+ F_{j}^{\rm Dir}(q^2) +
\frac{F_{j}^{\rm Pol}(0)}{1-q^2/m^2_j}~~,
\label{ff}
\end{equation}
where $m_1=m_{B^*}$, $m_0=m_{B^{**}}$ and we have marked the form
factors with a hat to stress that this formula does not hold in
the whole $q^2$ interval. As a matter of fact, as discussed above,
the $q^2$ range in which~(\ref{ff}) is expected to approximate
reliably the form factors is around the zero recoil point:
$q^2\simeq q^2_0\simeq m^2_B$. This follows from the  fact that
the model allows for a systematic derivative expansion, whose
first terms are represented by $\hat{F_j}$; terms of higher order
in the pion derivatives, which can be important at small $q^2$,
are suppressed for large $q^2$. This observation gives us a hint
to extrapolate to smaller values of $q^2$. Writing:
\begin{equation}
F_j(q^2)=\hat{F_j}(q^2) G_j(q^2)
\end{equation}
where  $j \in \{ 0,1\}$; this parameterisation has to satisfy
\begin{equation}
G_j(q_0^2)=1
\label{uno}
\end{equation}
as $q^2 \sim q_0^2$ is the region where $\hat{F_j}$ are a good
approximation of the form factors. Another condition that has to
be satisfied is the constraint:
\begin{equation}
\hat{F_1}(0) G_1(0)=\hat{F_0}(0) G_0(0)~, \label{due}
\end{equation}
which follows from  ${F_1}(0)={F_0}(0)$. It is reasonable to
assume that the corrections to $ G_j(q^2)\equiv 1$ arise from
terms with extra pion derivatives. Therefore we put
\begin{equation}
G_j(q^2)= 1 - \frac{E_\pi}{\alpha_j \Lambda_\chi}=1 -
\frac{(q_\pi\cdot p)}{\alpha_j m_B\Lambda_\chi}~,\label{gj}
\end{equation}
where $E_\pi$ is the pion energy in the $B$ rest frame,
$\Lambda_\chi= 1$ GeV and $\alpha_j  $ are free parameters. Since
(\ref{gj}) is equivalent to (under the assumption that $q_\pi^2 \ll
m_B^2$):
\begin{equation}
G_j(q^2)= 1 + \frac{q^2-m_B^2}{2 m_B \Lambda_\chi \alpha_j  }
\end{equation}
condition (\ref{uno}) is automatically satisfied. Formula
(\ref{due}) implies that $\alpha_0$ and $\alpha_1$ are related:
\begin{equation}
\frac{m_B}{2 \Lambda_\chi \alpha_1}=1- \left( 1
-\frac{m_B}{2\Lambda_\chi \alpha_0} \right)
\frac{\hat{F_0}(0)}{\hat{F_1}(0)}
\end{equation}
We could fix one of the two parameters from experimental data, were they
available. For the time being, in absence of such information, we have
to use some theoretical input. There exist many theoretical calculations
of the $B\to\pi$ couplings; for example quark models \cite{qmod,gns,lns}
predict $F_0(0)=0.20$ to $0.50$ with the exception of \cite{isgw} which gives a
very small value  $F_0(0)=0.09$. Chiral perturbation theory together with 
heavy quark effective theory gives 
$F_0(0)=0.38$ \cite{rep,semil}, QCD sum-rules give $F_0(0)=0.25$ to $0.40$
\cite{msuru,ballp,ballc}. Finally lattice results are $F_0(0)=0.27$ to $0.35$
\cite{ape,elc,ukqcd}.
We take as an input the result of the QCD sum rules calculation of 
\cite{ballp} that
gives $F_0(0)=0.30\pm 0.04$, in this way we obtain $\alpha_0=3.6$. It is
interesting to note that the rather large value of $\alpha_0$
obtained by this procedure indicates that the effective parameter
of the derivative expansion is not of the order of 1 GeV ($\simeq
\Lambda_\chi$), but larger, which means that, in spite of the fact
that this approximation should hold only at zero recoil point, it
gives reasonable estimates also at lower $q^2$. This conclusion is
corroborated by our estimate of the first correction to the
leading terms of the form factors, i.e. $F_j^{Dir}(q^2)$, which is
appreciable, but not very large ($10\%$ to $30\%$ of the total).

Letting the parameter $\alpha_0$ vary by $20\%$ allows to see how
the parameterisation affects the result.
In the Table the two form factors $F_1$ and $F_0$, including
the CQM correction, are given for few $q^2$ values near the maximum 
value $q_{max}^2=26.4$ GeV$^2$. The error refer a variation of the
values of $\alpha_0$ in the range $(2.9$--$4.3)$.

We have also reported results from other theoretical approaches. 
Let us note, as a concluding remark, that our calculation
includes, differently from other approaches based on the derivative
expansion, some deviations from the leading behaviour expected in the
soft pion limit. These extra terms, while sizeable,
are not such to change qualitatively the simple pole behaviour predicted
by the chiral effective theory.

\section*{Acknowledgements}

\noindent
A.D. acknowledges the support of the EC-TMR (European Community
Training and Mobility of Researchers) Program on ``Hadronic Physics with
High Energy Electromagnetic Probes''.

\section*{Figures}

\begin{figure}
\epsfxsize=7cm
\centerline{\epsffile{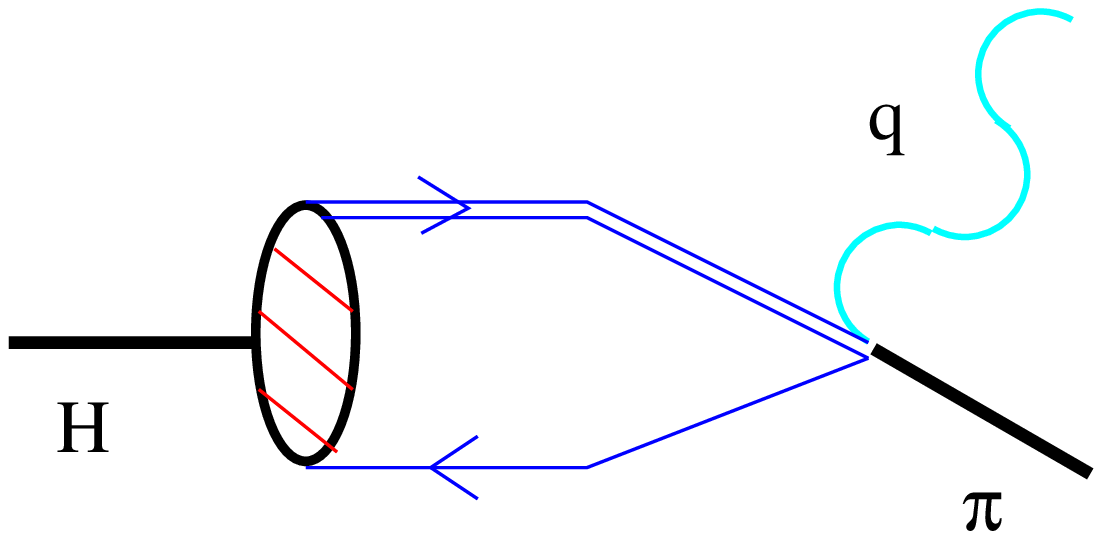}}
\noindent
{\bf Fig. 1} - {Diagram for the non-derivative contribution to the form factor
$B \to \pi$.}
\end{figure}

\begin{figure}
\epsfxsize=7cm
\centerline{\epsffile{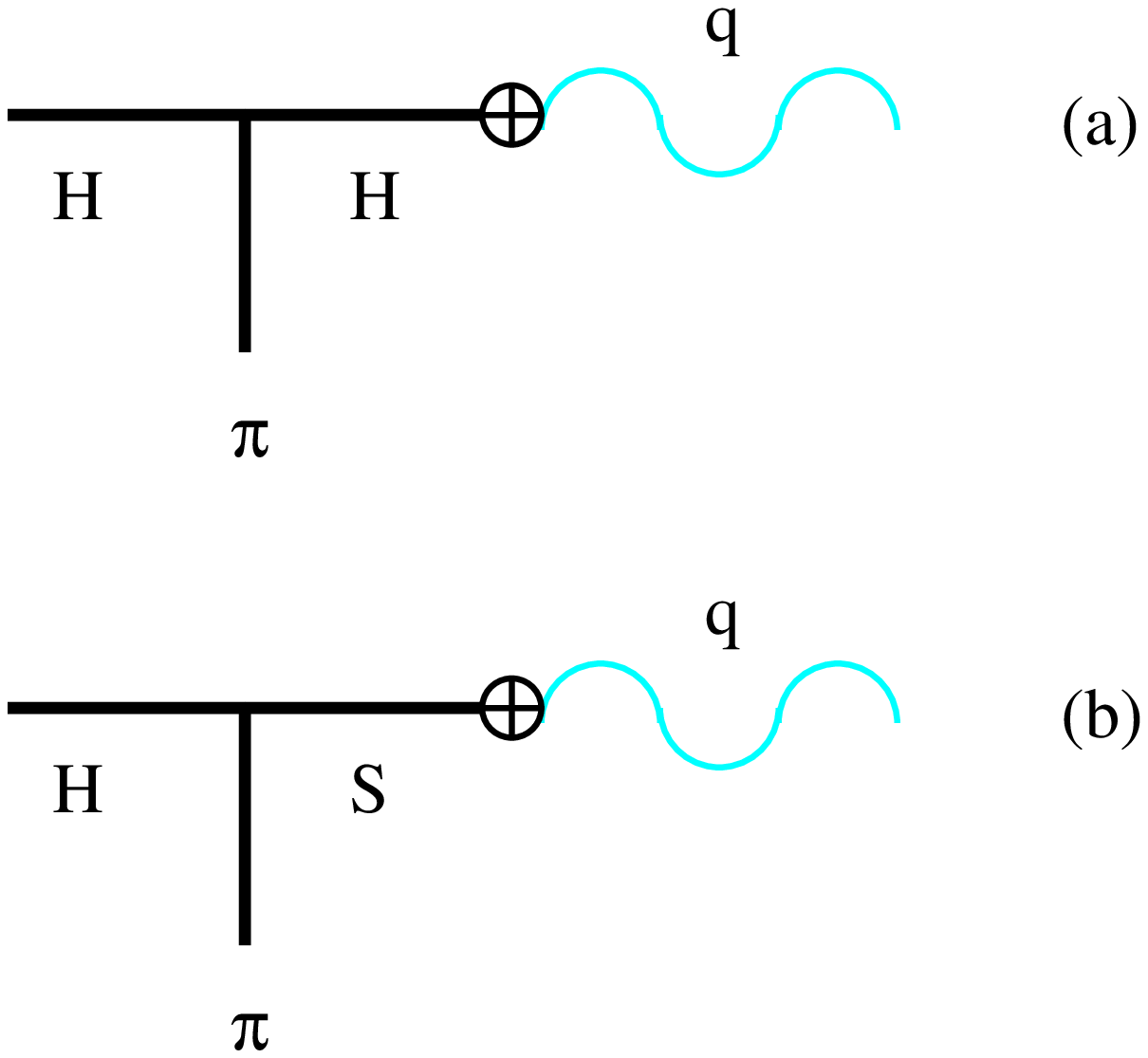}}
\noindent
{\bf Fig. 2} - {Diagram for the polar contribution to the form factor
$B \to \pi$.}
\end{figure}

\begin{figure}
\epsfxsize=7cm
\centerline{\epsffile{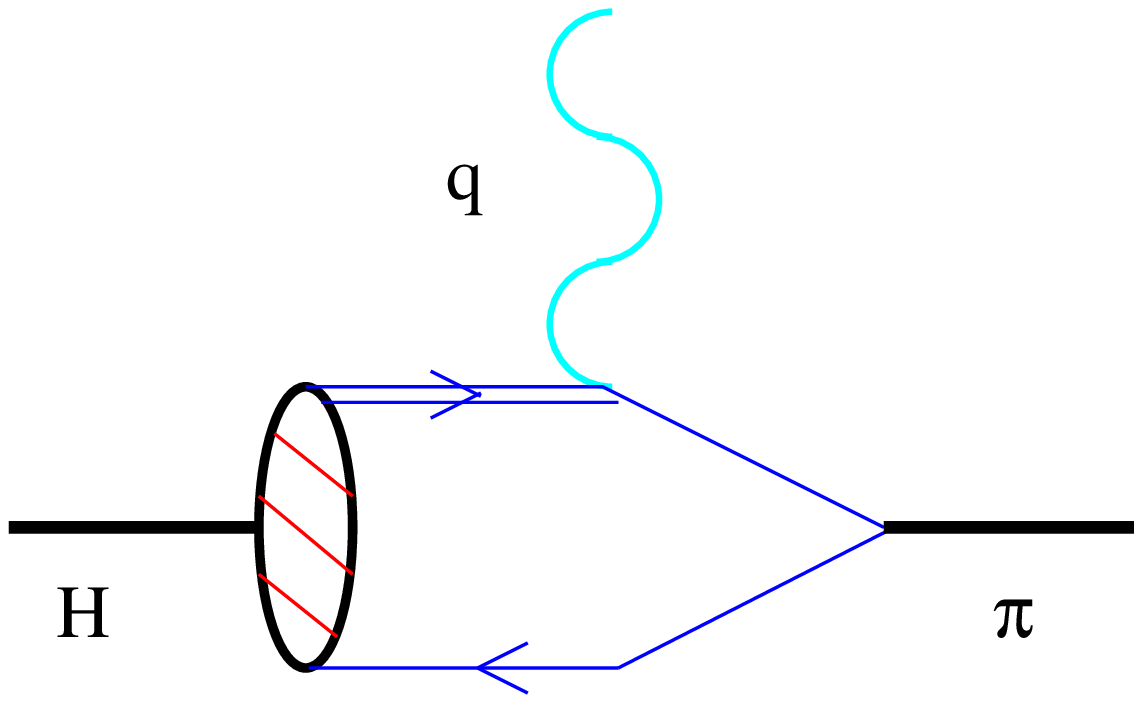}}
\noindent
{\bf Fig. 3} - {Diagram for the direct contribution to the form factor
$B \to \pi$.}
\end{figure}

\newpage
\onecolumn
\section*{Tables}

\begin{table}[h]
\begin{center}
\begin{tabular}{|c|c|c|c|c|}
\hline
$q^2$ & 14.9 GeV$^2$ & 17.2 GeV$^2$ & 20 GeV$^2$ & 26.4 GeV$^2$\\
\hline
CQM (this work)&&&&\\
\hline
$F_1^{B\pi}$& $1.58^{+0.28}_{-0.52}$& $2.06^{+0.27}_{-0.50}$&
$2.96^{+0.26}_{-0.47}$ &$13.78^{+0.13}_{-0.31}$\\
$F_0^{B\pi}$& $0.59^{+0.10}_{-0.18}$& $0.62^{+0.08}_{-0.14}$&
$0.65^{+0.05}_{-0.10}$ &$0.83 \pm 0.01$\\
\hline
IS (Quark Model) \cite{ist} &&&&\\
\hline
$F_1^{B\pi}$& 0.83& 0.96 & 1.19 & 3.14\\
$F_0^{B\pi}$& 0.48& 0.48 & 0.48 & 0.47\\
\hline
GNS (Quark Model) \cite{gns} &&&&\\
\hline
$F_1^{B\pi}$& 0.82& 1.05& 1.45&2.31\\
$F_0^{B\pi}$& 0.38& 0.40& 0.40&0.07\\
\hline
LNS (Quark Model) \cite{lns}&&&&\\
\hline
$F_1^{B\pi}$& 0.53& 0.57 & -- & --\\
$F_0^{B\pi}$& 0.69& 0.76 & -- & --\\
\hline
Ball (QCD light-cone) \cite{ballc}&&&&\\
\hline
$F_1^{B\pi}$&$0.85\pm 0.15$ & $1.1 \pm 0.2$& 1.6&--\\
$F_0^{B\pi}$& $0.5\pm0.1$& $0.55 \pm 0.15$& 0.7&--\\
\hline
Lattice (UKQCD) \cite{ukqcd}&&&&\\
\hline
$F_1^{B\pi}$&$0.85 \pm 0.20$ & $1.10 \pm 0.27$&$1.72 \pm 0.50$&--\\
$F_0^{B\pi}$& $0.46 \pm 0.10$& $0.49\pm 0.10$&$0.56\pm 0.12$&--\\
\hline
\end{tabular}
\caption{Form factors $F_1$ and $F_0$ at high $q^2$ values, near
($q^2_{max} \simeq 26.4 {\mathrm GeV}^2$) for $B \to \pi$ semi-leptonic
decays in CQM model and comparison with other calculations.
The error quoted for our result comes only from a $20\%$ variation in 
the parameter controlling the evolution from large $q^2$ values 
(where the calculation is more reliable) to smaller ones.}
\end{center}
\end{table}

\end{document}